\documentclass{article}
\usepackage{arxiv}
\usepackage[utf8]{inputenc} 
\usepackage[T1]{fontenc}    
\usepackage{doi}
\usepackage{breakurl}             
\usepackage{graphicx}
\usepackage{underscore}           
\usepackage{amsmath}
\usepackage{amssymb}
\usepackage{txfonts}
\usepackage{comment}
\usepackage{mathptmx}
\usepackage{xcolor}
\usepackage{array}
\usepackage{todonotes}
\usepackage{quiver}
\usepackage{ragged2e}
\usepackage{placeins}
\usepackage{algpseudocode}

\newcolumntype{C}[1]{>{\centering\let\newline\\\arraybackslash\hspace{0pt}}m{#1}}

\newcommand\imCMsym[4][\mathord]{%
  \DeclareFontFamily{U} {#2}{}
  \DeclareFontShape{U}{#2}{m}{n}{
    <-6> #25
    <6-7> #26
    <7-8> #27
    <8-9> #28
    <9-10> #29
    <10-12> #210
    <12-> #212}{}
  \DeclareSymbolFont{CM#2} {U} {#2}{m}{n}
  \DeclareMathSymbol{#4}{#1}{CM#2}{#3}
}
\newcommand\alsoimCMsym[4][\mathord]{\DeclareMathSymbol{#4}{#1}{CM#2}{#3}}

\imCMsym{cmmi}{124}{\CMjmath}
\imCMsym[\mathop]{cmsy}{113}{\CMamalg}
\imCMsym[\mathop]{cmex}{96}{\CMcoprod}
\alsoimCMsym[\mathop]{cmex}{97}{\CMbigcoprod}

\DeclareMathOperator{\Ob}{Ob}

\DeclareMathOperator{\Diag}{Diag}

\newcommand{\C}{{$\mathcal C$}}

\definecolor{darkred}{rgb}{0.55, 0.0, 0.0}

\begin{document}
\title{Compositional Exploration of Combinatorial Scientific Models}

%
%
\author{ \href{https://orcid.org/0000-0002-9374-9138}{\includegraphics[scale=0.06]{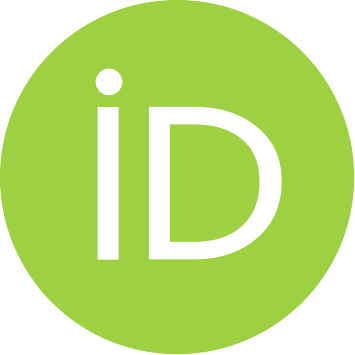}\hspace{1mm}Kristopher Brown} \\
	Department of Computer Science\\
	University of Florida\\
	\texttt{kristopher.brown@ufl.edu} \\
	\And
	Tyler Hanks \\
	University of Florida \\
	\texttt{thanks@ufl.edu} \\
	\And
	\href{https://orcid.org/0000-0002-1778-3350}{\includegraphics[scale=0.06]{orcid.pdf}\hspace{1mm}James Fairbanks} \\
	Department of Computer Science\\
	University of Florida\\
	\texttt{fairbanksj@ufl.edu} \\
}
\date{}
\renewcommand{\headeright}{}
\renewcommand{\undertitle}{}

\maketitle
\begin{abstract}
We implement a novel representation of model search spaces as diagrams over a category of models, where we have restricted attention to a broad class of models whose structure is presented by \C-sets. (Co)limits in these diagram categories allow the creation of composite model spaces from more primitive spaces. We present a novel implementation of the computer algebra of finitely presented categories and diagram categories (including their limits and colimits), which formalizes a notion of model space exploration. This is coupled with strategies to facilitate the selection of desired models from these model spaces. We demonstrate our framework by generating a tool which fits experimental data, searching an epidemiology-relevant subspace of mass-action kinetic models.

\end{abstract}

\section{Introduction}

Scientific progress is made through an iterative refinement of scientific theories and models. There is a reciprocal relationship between models and data: models are essential to the explanation and interpretation of data, while data is in turn essential to determining which model structure and parameters are appropriate to use. In this paper, we are concerned with the general process of selecting a new model in the face of data that has been collected and interpreted by one's current theory and hypothesized model. This process, while familiar to members across a wide range of scientific sub-disciplines, has not received attention at this general level, outside of philosophical treatments.

Category theory can be applied to this process at a very general level, providing insight and computational tooling to assist scientists across diverse fields in selecting new models in light of their data. A mathematical formalization of the model exploration process most immediately lends itself to building tools to partially automate this process, but it is also a worthwhile goal for the sake of scientific communication: presently, there is little standardization of how to present the procedure by which one selected a purported model, and scientists typically use informal descriptions or a program that performed the search. Better communication via a more structured artifact representing the search procedure could be important scientifically because, for example, statistical significance of results is contingent on the total number of models that were considered. Thus one application is to help quantify the phenomenon of procedural overfitting \cite{yarkoni2017choosing}.

Tools to organize and facilitate model exploration are of special importance recently due to the urgent need for reliable identification, analysis, and control of epidemic outbreaks. Epidemics can be shaped by a variety of factors; disease dynamics such as infection and recovery rates play a key role in determining whether a disease may become an epidemic. Concurrently, public policy adoptions such as quarantining, vaccination, and treatment can help to reduce the severity of an epidemic. Mathematical models are often utilized by epidemiologists to study epidemics and make policy suggestions. These interventions can be mathematically modeled as either changes to a model's parameters, or to a change in the model's structure. Thus this serves as a motivated and representative case study for the general phenomena of parameterized model exploration.

\paragraph{Structure of this paper} We first review some technical preliminaries of $\Diag(\mathcal{X})$, the category of diagrams in some category $\mathcal{X}$, as well as {\bf Petri}, a particular category of combinatorial models. We then demonstrate motivating examples of constructions of composite model spaces, with emphasis on the scenario of compartmental epidemiological models. Finally, we switch focus from model exploration to model selection: we discuss how the categorical structure of these model spaces facilitates selection and work through an extended epidemiological example. Note we represent composition of morphisms in diagrammatic order, i.e. $FG = G \circ F$.

\section{Model Spaces as Diagrams in Categories of Models}

\begin{figure}
    \centering
    \includegraphics[width=.9\textwidth]{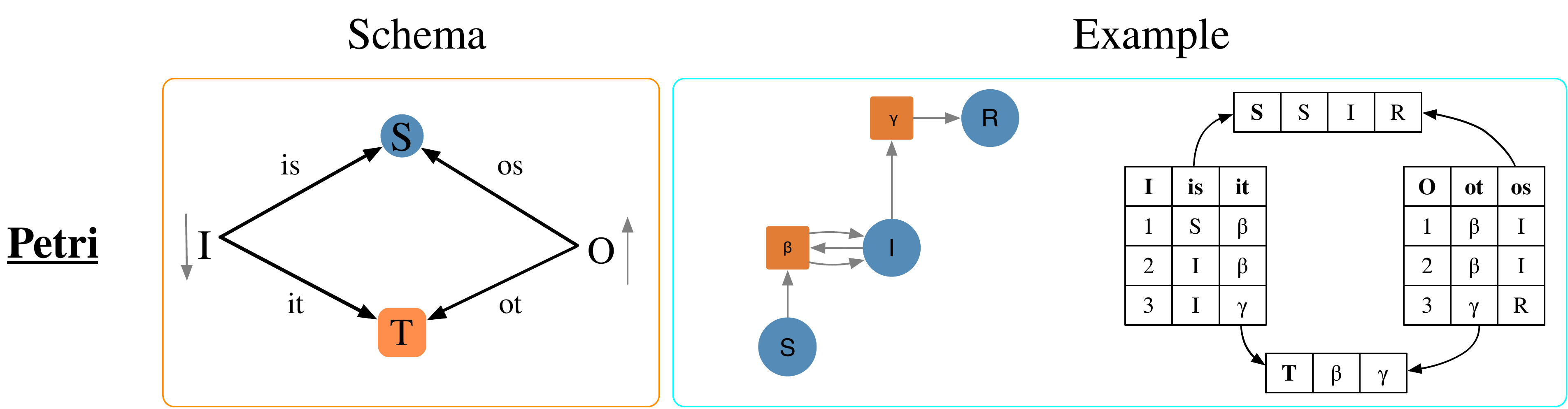}
    \caption{{\bf Petri} as a category of combinatorial models described by a schema with four objects and four arrows. With the SIR model as an example object of this category, we show how Petri nets can be depicted as a type of graph and as a kind of database.}
    \label{fig:cset}
\end{figure}

In order to avoid the philosophical issues of the ontology of models or the scope of the word `model'~\cite{frigg2020modelling}, we follow a tradition of thinking of models in terms of functors from syntax to semantics categories \cite{breiner2020categories}. For example, based on the grey-boxing functor in Pollard and Baez \cite{Baez2017}, we will interpret a Petri net as a dynamical system given by mass-action kinetics and parameterized by both the initial concentration for each state variable, and rate coefficient for each transition. This construction is a functor ${Params: Mono(Petri) \to Set}$ because it sends Petri nets to sets of parameters and Petri net monomorphisms to the inclusion of parameters via a pushforward. To further restrict our working definition of a model, we specifically consider copresheaves, or \C-sets, which are a categorical approach to relational databases \cite{dpo}. \C-{\bf Set} is bicomplete, which allows us to combine models via limits and colimits. We can view each \C-set as itself a category via the category of elements construction. This makes the class of scientific models, \C{\bf-Set} a category of categories, analogous to Halvorson and Tsementzis' characterization of the category of scientific theories \cite{halvorson2017categories}. Our paradigm example of a combinatorial structures is a whole-grain Petri Net, as seen in Figure~\ref{fig:cset}.

A diagram in a category $\mathcal{X}$ is a functor $D:J\rightarrow \mathcal{X}$ where $J$ is the indexing category. Thus a diagram can be thought of as a $J$ indexed family of objects in $C$. We sometimes implicitly regard an object of $\mathcal{X}$ as a diagram in $\mathcal{X}$ with shape {\bf 1}. Diagrams in a general {\it category} of models formalize a notion of {\it model spaces}. This is motivated by some practical features of how model spaces are used in practice: unlike a set of models, models are traversed with some notion of order. Furthermore, diagrams allow a distinction between the structure of one's model space from models themselves (e.g. we can talk about a grid-like structure of varying two hyperparameters without talking about the models themselves). Lastly, diagrams in a general category of models, rather than a preorder of models, allow us to express richer relationships between models than ``is prior to.''

We now define the category of (small) diagrams in some category of models, $\mathcal{X}$. Objects of $\Diag(\mathcal{X})$ are diagrams in $\mathcal{X}$, i.e. a tuple $(I, X)$ where $I$ is a small category in {\bf Cat}, called the {\it shape} of the diagram and $X: I\rightarrow\mathcal{X}$ is a functor in {\bf Cat}. $\Diag(Z)$ is a lax version of the slice category {\bf Cat}$/X$. A morphism between objects $(I,\ X)$ and $(J,\ Y)$ is a pair $(F,\ \phi)$, where $F: I\rightarrow J$ is a functor called the {\it shape map}. {\bf Cat}$/X$ demands $X=FY$, which is too strict for our purposes because it makes $\Diag(\mathcal{X})$ not bicomplete, even when $\mathcal{X}$ is bicomplete. In order to have access to (co)limits, we instead work with $\Diag(\mathcal{X})$, which only requires a natural transformation $\phi: X \rightarrow FY$, called the {\it diagram map}.

\begin{center}
$\begin{tikzcd}
	I && J \\
	& \mathcal{X}
	\arrow["F", from=1-1, to=1-3]
	\arrow[""{name=0, anchor=center, inner sep=0}, "X"', from=1-1, to=2-2]
	\arrow[""{name=1, anchor=center, inner sep=0}, "Y", from=1-3, to=2-2]
	\arrow["\phi", shorten <=6pt, shorten >=6pt, Rightarrow, from=0, to=1]
\end{tikzcd}$
\end{center}

We define {\it model exploration} as the process of building a space of possible models from simpler spaces. We distinguish this from {\it model selection}, which is the process of picking a model from the model space that addresses a practical problem \cite{ding2018model}. We can decouple these two processes; however, we recommend performing the exploration process with at least some considerations to facilitate selection. Notably, we suggest that the underlying graph of the diagram shape be either a directed acyclic graph (which can be topologically sorted) or a rooted graph, such that model selection has a designated starting point.

\section{Limits and colimits in diagram categories}
We review (co)limits in the category of diagrams, as described by Peschke and Tholen \cite{peschke2020diagrams}, where it is called $\Diag^\circ(\mathcal{X})$. We will present the constructions in more explicit detail and for arbitrary arities. Many constructions will implicitly range over $i \in I$ for some indexing set $I$.

The product in $\Diag(\mathcal{X})$, $(P,\ Z)\equiv {\prod}\ (A_i,\ X_i)$, is defined as a diagram whose shape is the product in {\bf Cat} of the diagram shapes, $\prod A_i$. $Z$ maps the object $\prod a_i \in \Ob P$ to the product in $\mathcal{X}$ given by $\prod X_i(a_i)$. Where $Z$ maps the morphism $f_i: \prod a_i \rightarrow \prod b_i$ is determined by the universal property of products in $\mathcal{X}$ of $P(\prod b_i)$. The projection map $(p_i,\ \pi_i): (P,\ Z)\rightarrow (A_i,\ X_i)$ is precisely the $i$th projection map of the shape-level product $P$ and the $\mathcal{X}$-level products from $Z$, respectively.

Equalizers in $\Diag(\mathcal{X})$ are likewise merely equalizers on the shape level and in $\mathcal{X}$. For a family of diagram morphisms $(F_i,\ \phi_i): (A,\ X)\rightarrow (B,\ Y)$, the equalizer $(E,\ \psi): (Eq,\ Z) \hookrightarrow (A,\ X)$ has $E$ given by the equalizer in {\bf Cat} of $F_i$. $\psi$ assigns to each object $a$ an inclusion $\iota_a: Eq(a) {\hookrightarrow} A(a)$ in $\mathcal{X}$ by taking the equalizers of morphisms $\phi_i(a)$ in $\mathcal{X}$. Naturality of $\psi$ forces $Z$ to map an object $a$ in $Eq$ to the domain of $\iota_a$, and $Z$ determines where to send a morphism $f: a \rightarrow a^\prime$ in $Eq$ by applying the universal property of $\iota_{a^\prime}$ to $\iota_aA(f)$.

Coproducts are also straightforward in $\Diag(\mathcal{X})$: $(C,\ Z)\equiv {\CMcoprod}\ (A_i,\ X_i)$. $C=\CMcoprod A_i$ is the coproduct at the shape level, and the shape maps $I_i$ of the inclusion morphisms $(I_i,\ \iota_i): (A_i,\ X_i)\hookrightarrow (C,\ Z)$, and the diagram data is copied over: $C(a)$ for some $a \in A_i$ is equal to $X_i(a)$ and a morphism from one of the included categories likewise gets mapped to where the original diagram mapped it. The inclusion diagram map for some $a \in A_i$ is the identity morphism in $\mathcal{X}$.

Coequalizers in $\Diag(\mathcal{X})$ require special attention. The detailed construction, found in \cite{peschke2020diagrams}, can also be expressed as a commutative diagram in $\Diag(\mathcal{X})$, shown in Figure~\ref{fig:diagcoeq}.

\begin{figure}[h!]
\centering
$
\begin{tikzcd}
	{(A,X)} & {(C,\ Lan_{FH}X)} \\
	{(B,Y)} & {(C,\ Lan_{H}Y)} \\
	& {(C,\ Z)}
	\arrow["{(F_,\ \phi_i)}"', shift right=3, from=1-1, to=2-1]
	\arrow["{(H,\ \kappa\gamma)}"', from=2-1, to=3-2]
	\arrow["{(H,\ \kappa)}", from=2-1, to=2-2]
	\arrow["{(id_C,\ \gamma)}", from=2-2, to=3-2]
	\arrow["{(id_C,\ \alpha_i)}", shift left=3, from=1-2, to=2-2]
	\arrow["{(FH,\ \lambda)}", shift left=1, from=1-1, to=1-2]
	\arrow[shift left=3, from=1-1, to=2-1]
	\arrow[from=1-1, to=2-1]
	\arrow[from=1-2, to=2-2]
	\arrow[shift right=3, from=1-2, to=2-2]
\end{tikzcd}$

\caption{Construction of a coequalizer $(H,\ \kappa\gamma)$ in $\Diag(\mathcal{X})$. Its shape map $H$ determined by the coequalizer of $F_i$. After computing the two left Kan extensions, $\alpha_i$ are determined by the universal property of $Lan_{FH}X$ in $[A, \mathcal{X}]$ applied to $\phi_i\kappa$, and $\gamma$ is their coequalizer in $[C, \mathcal{X}]$. Note that the data of a left Kan extension of a diagram in $\mathcal{X}$ is precisely a morphism in $\Diag(\mathcal{X})$ and that natural transformations in {\bf Cat} between functors into $\mathcal{X}$ can be represented as diagram morphisms with an identity shape map.}
    \label{fig:diagcoeq}
\end{figure}

\paragraph{Novel implementation}
For applied category theory to be computed, its concepts must be translated into a setting of finite presentations. To represent functors by a finite amount of data, their domains must be finitely presented, thus for diagrams we require every shape category to be presented by a finite reflexive graph, with vertices as object generators and edges as morphism generators, with morphisms implicitly quotiented by a finite list of equations between finite paths. Although mathematicians can quickly forget about the presentation as a mere aid to specifying a category, computational category theory requires taking these presentations of categories as its objects of study. This work adds functionality to Catlab.jl \cite{halter2020compositional}, a computational category theory library that permits the manipulation of \C-sets, finitely-presented categories, and diagrams.

Ideally the category of finitely presented categories, called {\bf CondGraph} by Borceux in \cite{borceux1994handbook} as they are graphs with sets of commutivity conditions, is closed under limits and colimits; this is almost true with one exception that will be mentioned. The construction of colimits stems from Borceux who constructs a left adjoint functor from {\bf CondGraph} into {\bf Cat}. Any colimits constructed in {\bf CondGraph} yield colimits in {\bf Cat} due to left adjoints preserving colimits. Products can also be generally computed as finitely presented categories, by a generalization of the process of computing finite presentations of product groups \cite{johnson1997presentations} and monoids \cite{howie1994constructions}. We restrict computing equalizers of finite functors $F_i: A \rightarrow B$ to cases where $B$ is a free category, such that the equalizer is a presented by a subgraph of the graph presenting $A$ along with the relevant subset of $A$'s path equations.

Beyond limits of finitely-presented categories, the implementation of limits and colimits in $\Diag(\mathcal{X})$ mostly follows as described above without special attention needed, as general limits and colimits of \C-sets are an existing Catlab feature. One exception is the computation of left Kan extensions, which proceeds as follows and is visualized in pseudocode below: we start with a diagram $X$ in \C{\bf-Set} with shape $J$ and a functor $F: J\rightarrow I$. The tensor-hom adjunction in {\bf Cat} establishes an equivalence between diagrams in {\bf Set} and diagrams in \C{\bf-Set}, which is applied in line 2 to $X$ to produce a corresponding $X^\prime$. Then, in line 3, the functor $- \times \mathcal{C}$ is applied to $F$ to produce a corresponding $F^\prime$ compatible with $X^\prime$. This data is in a form amenable to the pushout-based chase algorithm of Spivak and Wisnesky \cite{spivak2020fast}, which computes the left Kan extension of a diagram in {\bf Set}, shown in line 4. We can lastly reapply the tensor-hom equivalence to the resulting diagram in {\bf Set} and diagram morphism in {\bf Set}, in lines 5 and 6 respectively, to obtain the corresponding left Kan extension in \C{\bf-Set}.

\begin{algorithmic}[1]
\Function{leftkan_cset}{$X \color{gray}{::J \rightarrow \mathcal{C} \rightarrow \mathbf{Set}}$, $F\color{gray}{:: J \rightarrow I}$}
\State $X^\prime\color{gray}{:: J \times \mathcal{C} \rightarrow \mathbf{Set}}\ \ \  \color{black}{\longleftarrow\ \ \  \text{TensorHom}(X)}$
\State $F^\prime\color{gray}{:: J \times \mathcal{C} \rightarrow I \times \mathcal{C}} \color{black}\ \ \  \longleftarrow\ \ \  F \times \mathcal{C}$
\State $(Lan_{F^\prime}X^\prime\color{gray}{:: I \times \mathcal{C}\rightarrow \mathbf{Set}}\color{black},\ \alpha^\prime\color{gray}{:: X^\prime \rightarrow Lan_{F^\prime}X^\prime}\color{black}) \ \ \ \longleftarrow\ \ \  \text{LeftKan_Set}(X^\prime,\ F^\prime)$
\State $Lan_{F}X\color{gray}:: I \rightarrow \mathcal{C} \rightarrow \mathbf{Set}\color{black}\ \ \ \longleftarrow\ \ \  \text{TensorHom}^{-1}(Lan_{F^\prime}X^\prime)$
\State $\alpha\color{gray}:: X \rightarrow Lan_{F}X\color{black}\ \ \  \longleftarrow\ \ \  \text{TensorHom}^{-1}(\alpha^\prime)$
\State \Return $(Lan_FX,\ \alpha)$
\EndFunction
\end{algorithmic}

\paragraph{Examples}

\begin{figure}[h!]
    \centering
    \includegraphics[width=.9\textwidth]{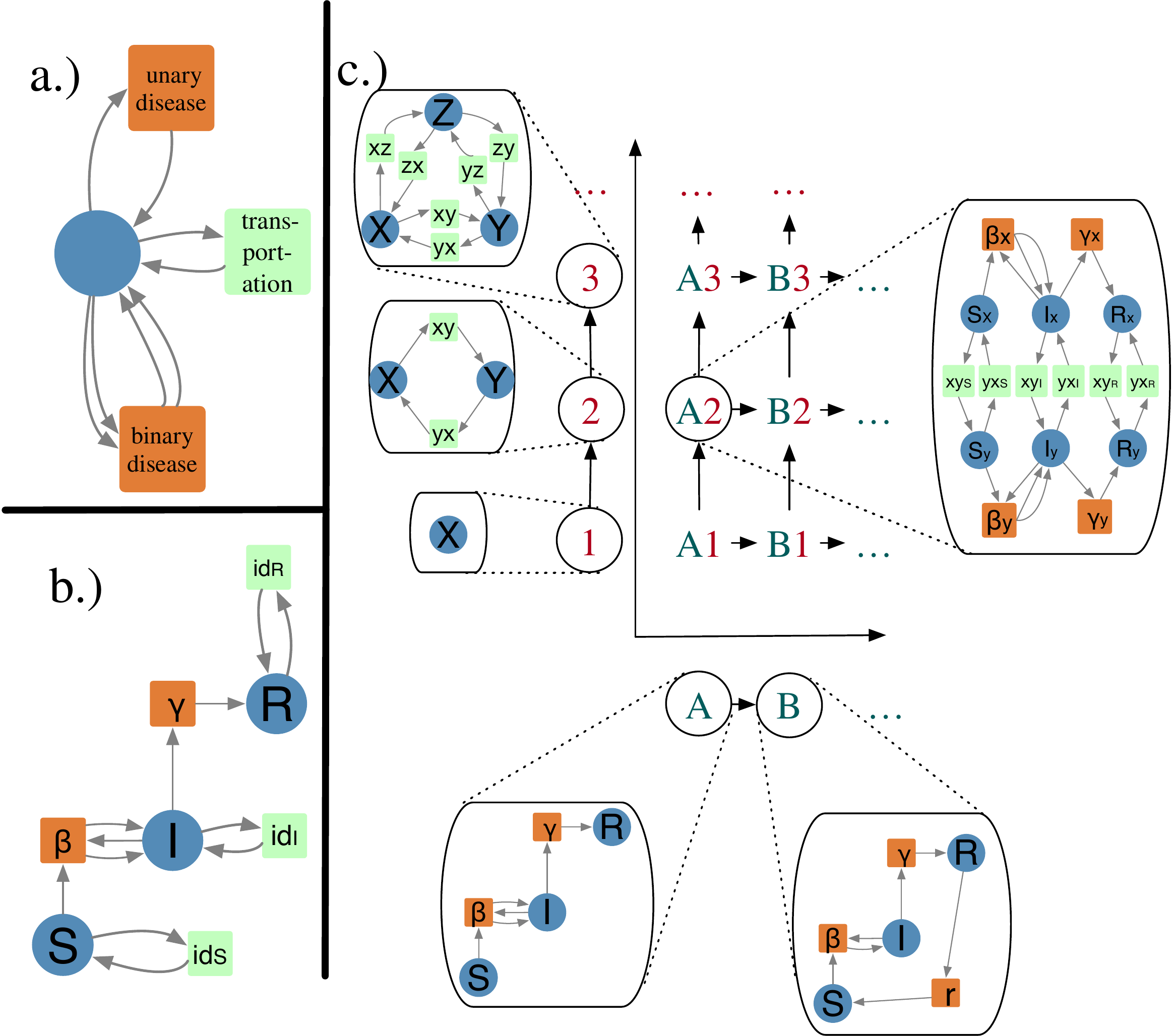}
    \caption{A discrete spatial stratification for a sequence of disease models, performed via pullback of path-shaped diagrams in {\bf Petri}. {\bf a.)} The apex of the cospan in $\Diag(\mathbf{Petri})$. Colors of boxes in subsequent panels indicate the data of the diagram maps into the apex of the cospan, i.e. homomorphisms into this Petri net. {\bf b.)} For ease of visualization, we implicitly elide the unary transitions corresponding to the other dimension of a stratification. This says, for example, the SIR model has a reflexive transportation transition for each state. This example is explicitly drawn out in this panel but remains implicit in the last panel for models \textcolor{teal}{$A$}, \textcolor{teal}{$B$}, \textcolor{darkred}{$1$}, \textcolor{darkred}{$2$}, and \textcolor{darkred}{$3$}.  {\bf c.)} A pullback of a geography dimension (one-, two, three-city models) and a disease dimension (SIR, SIRS, etc.). We visualize resulting grid-like shape of the pullback diagram and one of its underlying Petri nets.}
    \label{fig:pullbackex}
\end{figure}

\begin{figure}[h!]
    \centering
    \includegraphics[width=.9\textwidth]{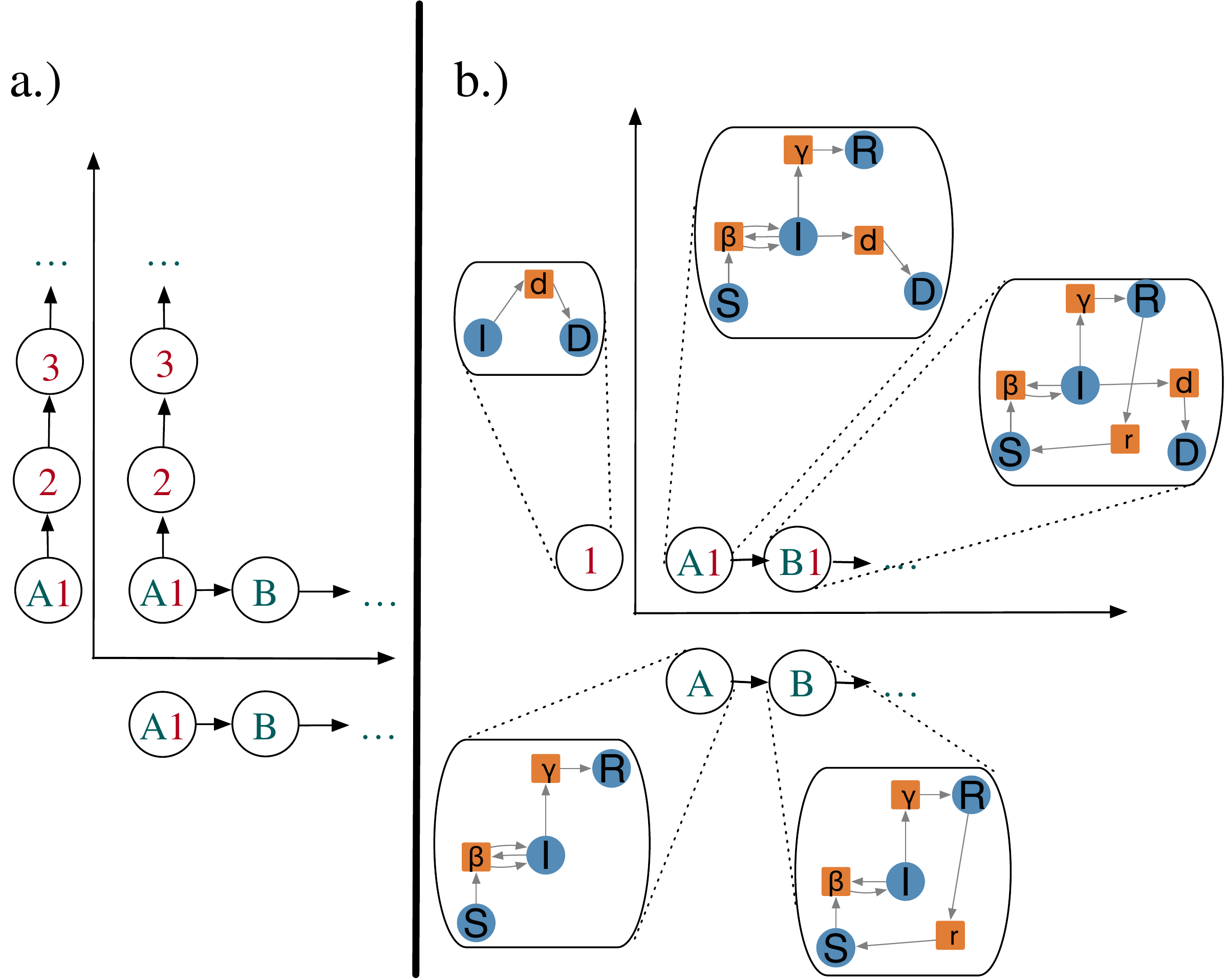}
    \caption{Two pushouts of spans of path-shaped diagrams in $\Diag(\mathbf {Petri})$. {\bf a.)} A diagram with a branching shape and no changes to the underlying Petri nets is created by pushing out the following span: the apex is the first element shared by both diagrams. {\bf b.)} A death transition is added to the sequence of disease models by computing the pushout of the a span with the single state $I$ as its apex. The shape maps out of this apex pick out \textcolor{teal}{$A$} and \textcolor{darkred}{$1$}, and the diagram maps pick out the $I$ state in each of their underlying Petri nets. In Figure~\ref{fig:wd} this operation is labeled as {\it Glue}.}
    \label{fig:pushoutex}
\end{figure}

We demonstrate the utility of diagram limits and colimits by showing epidemiologically-motivated examples of pullbacks and pushouts in the category of diagrams in {\bf Petri}. These specify spaces of mass-action kinetics models and are motivated by examples from Libkind et al. \cite{libkind2022algebraic}. One can gain intuition for how pullbacks work by considering the pullback of two path-shaped sequences of models, which performs model stratification. This a two-dimensional\footnote{Because pullbacks can be defined for an arbitrary arity, we can produce $n$-dimensional model spaces via pullbacks of $n$ path-shaped model spaces.} space of models, as shown in Figure~\ref{fig:pullbackex}. Alternatively, this is the product of diagrams in {\bf Petri}$/X$, where $X$ is given in panel {\bf a}. Pushouts are more intricate, reflecting the increased complexity of their construction. To highlight the versatility of this construction, we demonstrate a pushout at the `shape level' and at the `{\bf Petri} level' in Figure~\ref{fig:pushoutex}a and Figure~\ref{fig:pushoutex}b, respectively.

\section{Model exploration}

A paradigm case of model exploration in practice is model stratification. Stratification is characterized by taking some form of dynamics and reproducing it as {\it local} dynamics within multiple strata, which can interact in controlled ways. Epidemiological modelers in particular use this technique to examine effects of demographics or geography on the dynamics of disease.

We build upon the work in Libkind et al. \cite{libkind2022algebraic}, where the stratification of compartmental models in epidemiology is achieved via series of pullbacks in {\bf Petri}. Equivalently, these are products in the slice category {\bf Petri}$/X$. Pullbacks in $\Diag(\mathbf{Petri})$ formalize the informal structure implicit in taking all possible pairs of pullbacks from a set of disease dynamics models and a set of stratification models. Our contribution is a level shift from pullbacks in {\bf Petri} to pullbacks in $\Diag(\mathbf{Petri})$: limits in {\bf Petri} are combinations of models, while limits in $\Diag(\mathbf{Petri})$ are combinations of model spaces.

\begin{figure}[h!]
    \centering
    \includegraphics[width=.8\textwidth]{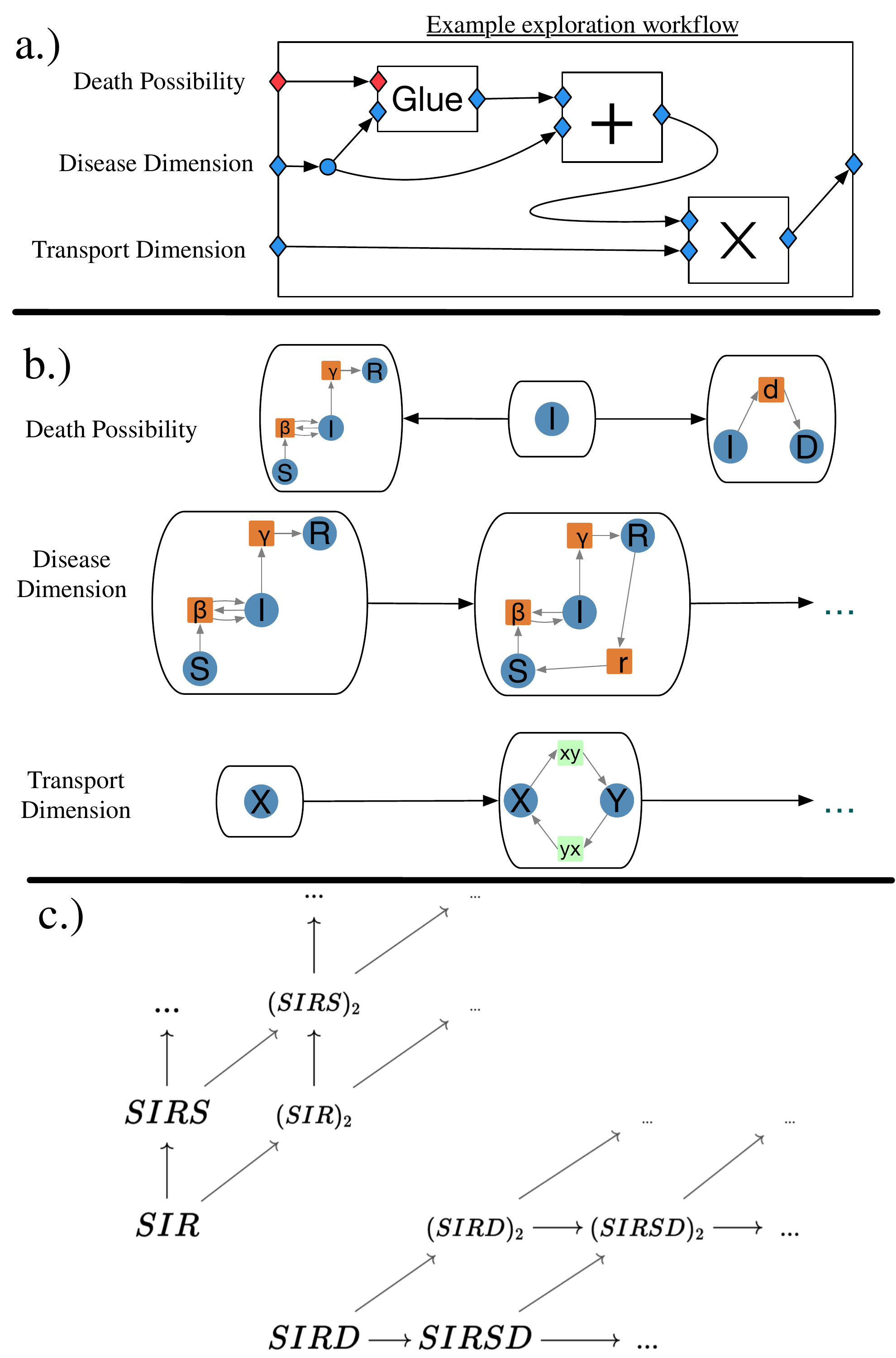}
    \caption{ {\bf a.)} A model exploration workflow represented as a typed wiring diagram. Models are combined by gluing, colimits, and limits. \textcolor{blue}{Diagrams in {\bf Petri}$/X$} are blue, and \textcolor{red}{spans in {\bf Petri}} are red. {\bf b.)} Example three inputs fed into this workflow {\bf c.)} Overall resulting composite model space. }
\label{fig:wd}
\end{figure}

Our example of one possible exploration of a space in {\bf Petri} combines the three (co)limit examples seen so far into one high-level operation, visualized in Figure~\ref{fig:wd}a. This workflow performs a stratification defined by a product in $\Diag(\mathbf{Petri}/X)$. The example also embellishes the disease dynamics dimension with an alternative that includes a death transition, i.e. the result of Figure~\ref{fig:pushoutex}b. However, that pushout effectively {\it replaces} the original disease dynamics, whereas this example adds the embellished version as an additional option via a coproduct in $\Diag(\mathbf{Petri}/X)$. The resulting space is depicted in Figure~\ref{fig:wd}b.

The three high-level operations of this example could be presented as basic recipes for a practicing scientist to work with. However, these and other useful abstractions can also be built compositionally from low level operations, such as primitive constructors for diagrams, diagram morphisms, and their (co)limits, or from intermediate-level abstractions. This composition is formalized by the operad of directed wiring diagrams \cite{patterson2021wiring}. By considering a slice category of directed wiring diagrams that encodes the basic argument types and signatures of primitive functions, we obtain a basic notion of type checking in order to prevent some user errors \footnote{Because this workflow type system does not involve full dependent types, an error such as taking the pushout of two morphisms which do not share a common domain would be a runtime error.}. This formal diagrammatic syntax  provides a visualizable and hierarchical domain specific language for model exploration. In addition to the clarity that comes from visualization, this syntax explicitly separates the workflow from the arguments it is applied to at runtime, allowing the application of same workflows to varying initial model spaces. Furthermore, isolating this wiring diagram syntax from the model space composition semantics allows us to decouple the  model space exploration from its implementation details, as another runtime decision could be whether to eagerly evaluate the composite model space all at once or lazily unfold it, the latter of which is necessary if one wanted to compute with infinite model spaces.


\section{Model selection}

Our paradigm case of model selection in Diag(\textbf{Petri}) is choosing a Petri net model which achieves a best-fit parameterization for a given data set which contains time series data for some set of species. Petri nets are given semantics of systems of ordinary differential equations by applying the law of mass action, as described by
Baez and Pollard \cite{Baez2017}. Then, the {\it loss} for a given Petri net with states $S$ and transitions $T$ (with a parameter vector $\vec p \in \mathbb{R}_{>0}^{|S|+|T|}$, representing the initial values for each state and the rates for each transition) is given in Equation \ref{eq:fit}, where $y$ represents data we wish to fit (a trajectory collected within the SIR model paradigm) and $\hat y$ represents the SIR trajectory as predicted by the Petri net, parameterized by the current parameters and time.

\begin{equation}
  \label{eq:fit}
  \begin{gathered}
    Loss(y,\hat y,\vec p) = \sum_{t \in T} \sum_{s \in \{S,I,R\}} (y_s(t) - \hat y_s(\hat p,\ t))^2
  \end{gathered}
\end{equation}

One key goal of formalizing model spaces as diagrams was to provide enough structure to determine an order for possible models to be evaluated, which is crucial when it is infeasible or impossible to evaluate every model in one's model space. When the underlying graph of the diagram's shape is acyclic or has a designated root object, breadth-first search is a natural choice of ordering. Given a category of models \C~and a model evaluation function $P:\Ob$ \C~ $\rightarrow \mathbb{R}_{>0}^{op}$, where a low evaluation score is preferred such as $P$ assigning the minimum loss over all parameter assignments, we define these to be {\it compatible} when $P$ is a functor. This means our models monotonically improve with respect to our goal as we traverse the model space along its morphisms. Notice that if a $P\hookrightarrow Q$ in ${\bf Petri}$ there is always a parameter choice of $Q$ that recovers the loss of the best parameterization of $P$ by setting all of the concentrations/rates for states/transitions in $Q$ that are not in $P$ to 0. Thus a monomorphism of Petri nets can only decrease the optimal loss.


\begin{figure}[h!]
    \centering
    \includegraphics[width=1.05\textwidth]{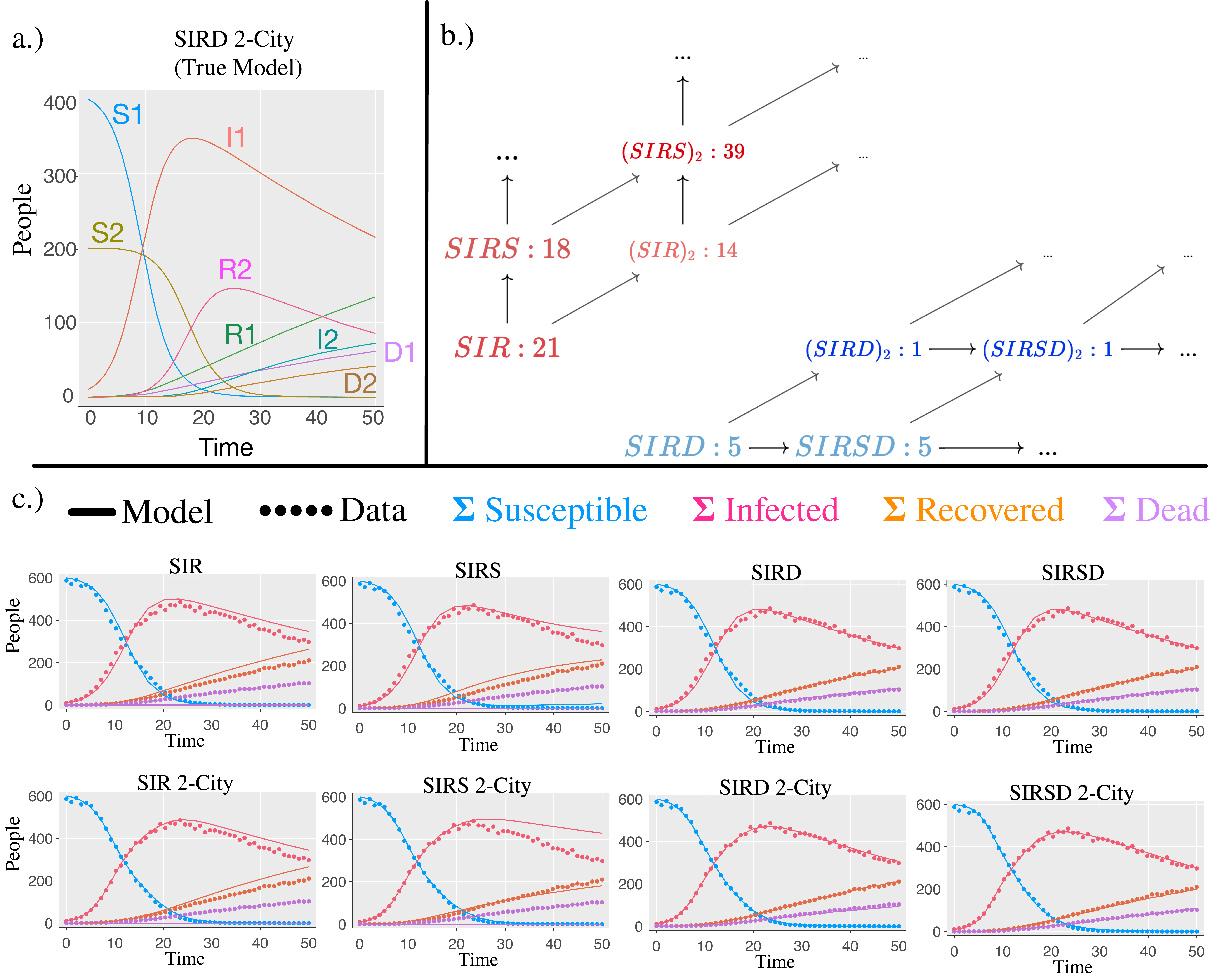}
    \caption{{\bf a.)} $(SIRD)_2$ trajectories which are taken as a ground-truth model {\bf b.)}   Annotation of the model space from Figure~\ref{fig:wd} with the loss scores associated with trying to fit data generated by a $(SIRD)_2$ model, normalized to the $(SIRD)_2$ model having a loss of $1.0$. {\bf c.)} SIRD trajectories (where $S$ is the sum of all $S$-associated states of the model, for example) that best match the fitted data.}
    \label{fig:results}
\end{figure}

There are many more ways in which the diagram structure can be leveraged during model selection. Firstly, beyond breadth-first search, there is great flexibility in dynamically selecting a traversal order based on the evaluation of models seen so far. For example, a discretized gradient computation of a product model space such as in Figure~\ref{fig:pullbackex} provides a data-informed guess for which directions are most promising to explore. We can also leverage the rich structure of diagrams constructed via limits and colimits to propagate prior information from our simpler model spaces into composite model spaces. For example, probability distributions on the states of the two-city model and the SIR model can be combined to give a probability distribution on the states of their combined model (shown in Figure~\ref{fig:pullbackex}): this formalizes the intuition that, if 90\% of the population is initially susceptible (to the best of our knowledge) and 33\% of the population initially is in City X, a reasonable guess for the initial percentage of susceptible people in City X is $90\% \times 33\% = 30\%$. These informed initial guesses are essential to the practical computation of best fit parameterizations. Furthermore, this limit and colimit information is  essential for our loss function in Equation \ref{eq:fit}, which requires a model to provide predictions commensurate with an SIR data set, even if the model contains dozens of states, none of which correspond directly to S, I, nor R. The projection map into the disease dimension allows us to `integrate' the states of a composite model along this axis.

This data fitting problem is compatible with the category of monomorphisms of {\bf Petri}$/X$. Given a best fit of some model $A$ to a data set, we know that the best fit of some model $B$ related by a morphism $A\hookrightarrow B$ is at least as good. Moreover, the data of the homomorphism tells us how to choose parameters for $B$ such that equal performance to $A$ is achieved, by setting newly introduced parameters to zero. This is one virtue of considering diagrams in a general {\it category} of models in contrast to a preorder of models. This approach can be used to produce an initial choice of parameters for optimization when we do not have prior distributions.

Taking all these strategies together, we demonstrated the feasibility of our approach by performing model selection for a best fit model on the search space described in Figure~\ref{fig:wd}c. To generate sample data, we simulated the $(SIRD)_2$ model from the space with a reasonable choice of parameters for 50 time steps. We then sampled the resultant trajectories at 50 evenly spaced points and added a small amount of Gaussian noise. The DiffEqFlux.jl package \cite{rackauckas2019diffeqflux} and the loss function in Equation \ref{eq:fit} were used to perform parameter fitting for the models in the search space. The results in Figure~\ref{fig:results} show each of the eight model's performance against the sample data, crucially showing that the true model $(SIRD)_2$ and $(SIRSD)_2$, achieved the lowest loss by a significant amount.

Because of the compatibility between the structure of our category of models and loss function, there are two observations we can make about the evaluated models, visualized in Figure~\ref{fig:results}b. Firstly, because there is a morphism $(SIRD)_2 \hookrightarrow (SIRSD)_2$, it is reasonable that $(SIRSD)_2$ is able to fit the experimental data equally well (i.e., by setting the reinfection rate to be zero). However, $(SIRD)_2$ can be identified as the simplest model that accounts for the data, where this measure of simplicity comes from the structure of the shape category. We also observe instances loss {\it increasing} along the morphisms into the $SIRS$ two-city model. Because the loss is guaranteed to decrease across morphisms, as it was formally defined, we can conclude this increase is purely an artifact of the numerical challenges of multivariate nonlinear optimization in a finite amount of time.

\section{Generalizations and future work}

\paragraph{Generalization 1: Keep model syntax fixed, change model semantics}
    Mass-action kinetics is a reasonable approximation in many chemical domains; however, biological processes are sometimes best modeled by other functional forms. For example, in the exploration of models of cancer treatment, cancer growth rate can be modeled as exponential ($\dot N = rN$), logistic ($\dot N = rN(1-\frac{N}{k})$),
    von Bertalanffy ($\dot N = \alpha N^\lambda - \beta N^\mu)$, or Gompertz ($\dot N = rN$ln$\frac{k}{N}$) functional forms. Likewise, there are alternate models for death rates due to chemotherapy, such as the log-kill hypothesis ($\dot N = -c(t)N$),  Norton-Simons hypothesis ($\dot N = -c(t)f(N)$), or the $\epsilon$-max model ($\dot N = \frac{-c(t)N}{N + \delta}$). Given the multiplicity of possible cancer models and treatment options, a similar Petri net exploration to Figure~\ref{fig:wd} could be used to structure the problem of finding an optimal treatment; however, depending on whether boxes are tagged as having a certain functional form via slicing over a Petri net with multiple transition types, the {\it semantics} of these Petri nets will be evaluated differently when comparing to experimental data.

\begin{figure}[h!]
    \centering
    \includegraphics[width=.6\textwidth]{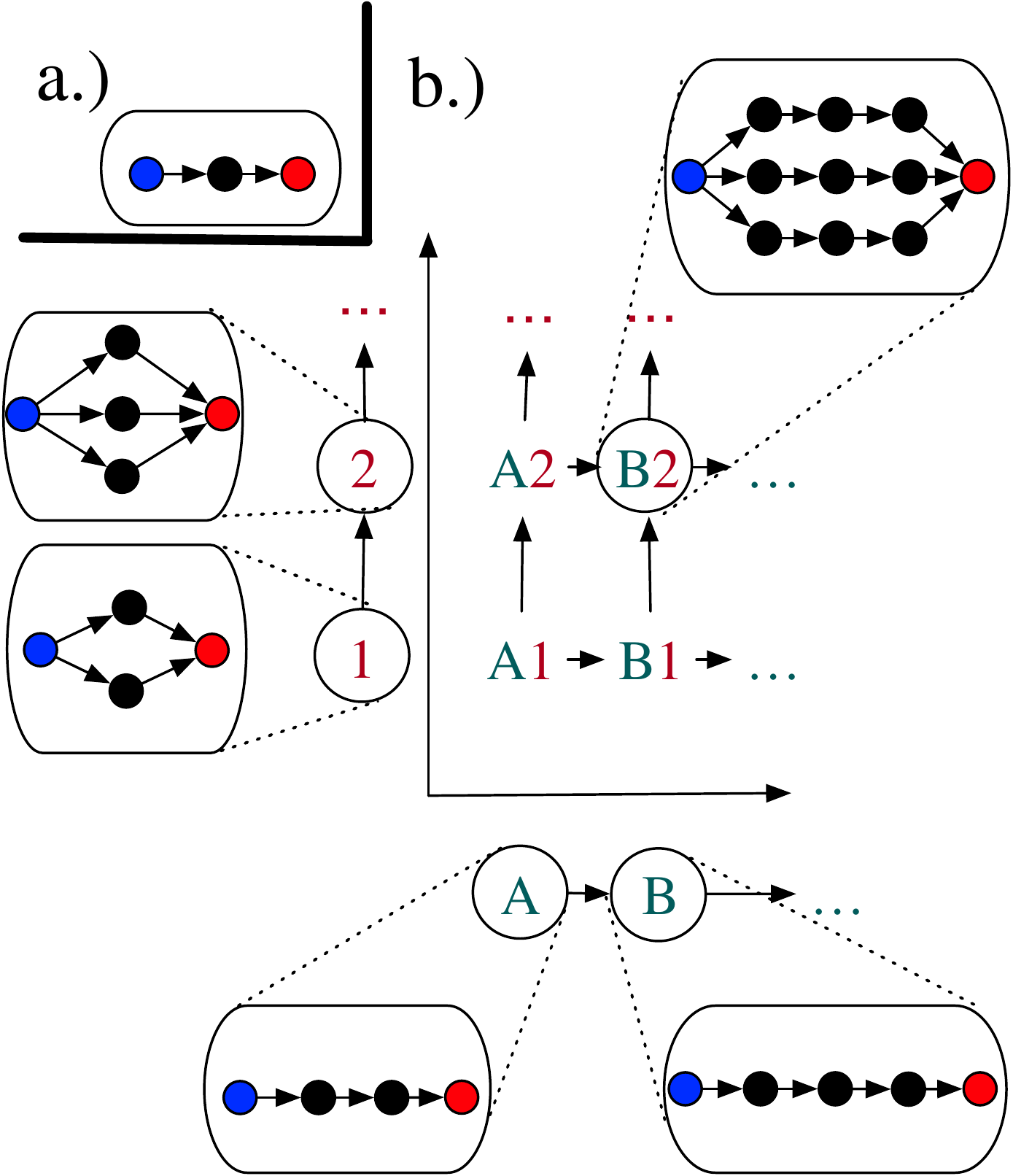}
    \caption{Neural network search space encoded as the product of two diagrams in \textbf{ReflexiveGraph}/$X$ (reflexive edges are suppressed for visualization). Vertices represent tensors, while edges represent layer operations. Blue and red represent input and output nodes, respectively, while black represents hidden nodes. {\bf a.)} a depiction of $X$, the base point of the slice, which enforces the single source/single sink requirement. {\bf b.)} The composite search space. Moving along the vertical axis widens the network while traversing the horizontal axis deepens it.}
    \label{fig:NN}
\end{figure}

\paragraph{Generalization 2: Change model syntax and semantics}
Up until this point, we have only considered {\bf Petri}$/X$ as a source of models with combinatorial structure. However, our framework is general to diagrams in any bicomplete category, for example \C{\bf-Set}$/X$ with any choice of finitely-presented \C. One combinatorial model exploration problem with a rich literature is neural architecture search, which involves finding a neural network architecture that achieves desired accuracy on a given problem. Many techniques model the search space as a directed graph with operations such as convolution and pooling on edges and data such as tensors on nodes. For example, Liu et al. represent architectures as single source/single sink directed acyclic graphs over a small set of primitive operations \cite{LiuNAS}. We represent this by working in \textbf{ReflexiveGraph}/$X$, the category of reflexive graphs and graph homomorphisms sliced over the graph in Figure~\ref{fig:NN}a, which enforces the single source/single sink requirement. We can then encode two linear diagrams over this category: one to represent an iterative deepening of the architecture and another to represent an iterative widening. These two diagrams can be combined via a product in the category of diagram to produce the search space in Figure~\ref{fig:NN}b.

The evaluation criteria of finding an optimal parameterization relative to a data set is valuable in this setting, much like the case of Petri nets. As the structure of this category of models is compatible with this parameter estimation problem, this means morphisms between models can aid in the search process and parameter initialization.

\paragraph{Alternative selection function: lowering the curve}
An example of a valid loss function that is {\it incompatible} with the structure of {\bf Petri} is one where we seek to optimize parameters to minimize the maximum infected population concentration over a fixed time interval, as formalized in Equation \ref{eq:lower}. Here the model could also be judged by how low this value is, possibly in conjunction with other desiderata (for example, some policies may not be as politically feasible). In this case, it will not be the case that a morphism in {\bf Petri} corresponds to an improvement in lowering the curve. One consequence of this fact is a lack of regularity of the loss landscape. Although gradient-like heuristics can still be used, there is no longer clear criteria for when to terminate the search in the case of infinite model spaces.

\begin{equation}
  \label{eq:lower}
  \begin{gathered}
    Loss\prime(\hat y,\vec p) = \underset{t \in T}{max}\  \hat y_I(\hat p, t)
  \end{gathered}
\end{equation}

\paragraph{Future work}
Many elements of future work are required to develop our prototype into a full-fledged software tool for scientists with no computer science or category theory background to use. A GUI using the wiring diagram formalism for constructing workflows would improve accessibility. Another task is to develop a high-level library on top of the primitive abstractions of limits and colimits of diagrams to capture common design patterns used by scientists in practice. This could also involve introducing new primitives for producing model spaces: for example, by defining a set of open rewrite rules, one induces a diagram in the double category of structured cospan rewrites \cite{cicala2019rewriting}. The development of a lazy implementation of limits and colimits would allow computation with infinite model spaces. Another future research direction involves models which are themselves hierarchical: multiple loss functions could be used in parallel to guide the selection process, biasing search towards models where both individual components perform well in addition to cohering with each other to achieve an overall task.

\section{Conclusion}

Considering diagrams in a category of models as model spaces allows us to formalize model exploration and model selection, aspects of iterative scientific practice that are usually treated with an ad hoc paradigm or left informal. In particular, the notions of pullbacks and pushouts in this category of model spaces captures the commonplace notions of multidimensional search (model space stratification) and branching possibilities. Limits and colimits in $\Diag(\mathcal{X})$ are a precise and compositional language to specify the hierarchical construction of interesting topologies of model spaces.

We presented one mechanism by which category theory can be applied to scientific computing to improve the transparency of its methods as well as its flexibility and robustness to the fast-changing goals of scientists. We built a very general tool for scientific modeling and applied it to solve small examples of real problems. Implementing this required the first practical implementations of limits and colimits for categories of diagrams in \C{\bf-Set}. \C{\bf-Set} is a well-behaved category with straightforwardly computable (co)limits, and $\Diag($\C{\bf-Set}$)$ inherits this behavior. We demonstrated building up model exploration workflows, with example applications to Petri nets and neural network architectures. We lastly showcased an example model selection framework for finding parameterized epidemiological models that can deduce the simplest model consistent with an experimentally observed trajectory as well as incorporate prior knowledge about stratification dimensions.

\paragraph{Acknowledgements}
The authors would like to thank Evan Patterson, David Spivak, and Tim Hosgood for their helpful conversations.

\paragraph{Funding} The authors were supported by DARPA Award W911NF2110323.

\FloatBarrier

\bibliographystyle{splncs04}
\bibliography{references}

\begin{thebibliography}{10}
\providecommand{\url}[1]{\texttt{#1}}
\providecommand{\urlprefix}{URL }
\providecommand{\doi}[1]{https://doi.org/#1}

\bibitem{Baez2017}
Baez, J.C., Pollard, B.S.: A compositional framework for reaction networks.
  Reviews in Mathematical Physics  \textbf{29}(09),  1750028 (sep 2017).
  \doi{10.1142/s0129055x17500283},
  \url{https://doi.org/10.1142%2Fs0129055x17500283}

\bibitem{borceux1994handbook}
Borceux, F.: Handbook of categorical algebra: volume 1, Basic category theory,
  vol.~1. Cambridge University Press (1994)

\bibitem{breiner2020categories}
Breiner, S., Denno, P., Subrahmanian, E.: Categories for planning and
  scheduling. Notices of the American Mathematical Society  \textbf{67}(11)
  (2020)

\bibitem{dpo}
Brown, K., Hanks, T., Patterson, E., Fairbanks, J.P.: Computational
  category-theoretic rewriting. CoRR  \textbf{abs/2111.03784} (2021),
  \url{https://arxiv.org/abs/2111.03784}

\bibitem{cicala2019rewriting}
Cicala, D.: Rewriting structured cospans: A syntax for open systems. University
  of California, Riverside (2019)

\bibitem{ding2018model}
Ding, J., Tarokh, V., Yang, Y.: Model selection techniques: An overview. IEEE
  Signal Processing Magazine  \textbf{35}(6),  16--34 (2018)

\bibitem{frigg2020modelling}
Frigg, R., Nguyen, J.: Modelling nature: An opinionated introduction to
  scientific representation. Springer (2020)

\bibitem{halter2020compositional}
Halter, M., Patterson, E., Baas, A., Fairbanks, J.: Compositional scientific
  computing with catlab and semanticmodels. arXiv preprint arXiv:2005.04831
  (2020)

\bibitem{halvorson2017categories}
Halvorson, H., Tsementzis, D.: Categories of scientific theories. Categories
  for the Working Philosopher pp. 402--429 (2017)

\bibitem{howie1994constructions}
Howie, J.M., Ru{\v{s}}kuc, N.: Constructions and presentations for monoids.
  Communications in Algebra  \textbf{22}(15),  6209--6224 (1994)

\bibitem{johnson1997presentations}
Johnson, D.L., et~al.: Presentations of groups. No.~15, Cambridge university
  press (1997)

\bibitem{libkind2022algebraic}
Libkind, S., Baas, A., Halter, M., Patterson, E., Fairbanks, J.: An algebraic
  framework for structured epidemic modeling. arXiv preprint arXiv:2203.16345
  (2022)

\bibitem{LiuNAS}
Liu, H., Simonyan, K., Vinyals, O., Fernando, C., Kavukcuoglu, K.: Hierarchical
  representations for efficient architecture search (2017).
  \doi{10.48550/ARXIV.1711.00436}, \url{https://arxiv.org/abs/1711.00436}

\bibitem{spivak2020fast}
Meyers, J., Spivak, D.I., Wisnesky, R.: Fast left-kan extensions using the
  chase. Journal of Automated Reasoning (to appear)  (2022).
  \doi{10.48550/ARXIV.2205.02425}

\bibitem{patterson2021wiring}
Patterson, E., Spivak, D.I., Vagner, D.: Wiring diagrams as normal forms for
  computing in symmetric monoidal categories. arXiv preprint arXiv:2101.12046
  (2021)

\bibitem{peschke2020diagrams}
Peschke, G., Tholen, W.: Diagrams, fibrations, and the decomposition of
  colimits. arXiv preprint arXiv:2006.10890  (2020)

\bibitem{rackauckas2019diffeqflux}
Rackauckas, C., Innes, M., Ma, Y., Bettencourt, J., White, L., Dixit, V.:
  Diffeqflux.jl-a julia library for neural differential equations. arXiv
  preprint arXiv:1902.02376  (2019)

\bibitem{yarkoni2017choosing}
Yarkoni, T., Westfall, J.: Choosing prediction over explanation in psychology:
  Lessons from machine learning. Perspectives on Psychological Science
  \textbf{12}(6),  1100--1122 (2017)

\end{thebibliography}
\end{document}